\documentclass[10pt]{article}
\usepackage{opex3}
\usepackage{amsmath}    
\usepackage{amsfonts}
\usepackage{bm}
\usepackage{amssymb}
\usepackage{appendix}
\usepackage{graphicx}   
\usepackage{subfigure}  
\usepackage{comment}
\usepackage{cite}
\usepackage{color}
\usepackage{url}
\newcommand{\re}{\textrm{Re}}
\newcommand{\im}{\textrm{Im}}

\begin{document}
\title{Controlling mode competition by tailoring the spatial pump distribution in a laser: A resonance-based approach}

\author{Alexander Cerjan,$^{1,*}$ Brandon Redding,$^{2}$ Li Ge,$^{3,4}$ Seng Fatt Liew,$^{5}$ Hui Cao,$^{5}$ and A.~Douglas Stone$^{5}$}
\address{$^{1}$Department of Electrical Engineering, and Ginzton Laboratory, Stanford  University,  Stanford,  California  94305,  USA \\
$^{2}$Naval Research Laboratory, 4555 Overlook Ave SW, Washington DC 20375 USA \\
$^{3}$Department of Engineering Science and Physics, College of Staten Island, CUNY, Staten Island, NY 10314, USA \\ 
$^{4}$The Graduate Center, CUNY, New York, NY 10016, USA\\
$^{5}$Department of Applied Physics, Yale University, New Haven, Connecticut 06520, USA}
\email{$^{*}$acerjan@stanford.edu}

\date{\today}

\begin{abstract}
We introduce a simplified version of the steady-state \textit{ab initio} laser theory
for calculating the effects of mode competition
in continuous wave lasers using the passive cavity resonances. 
This new theory harnesses widely available numerical methods 
that can efficiently calculate the passive cavity resonances, with
negligible additional computational overhead.
Using this theory, we demonstrate that the pump profile of the laser cavity
can be optimized both for highly multi-mode and single-mode emission. An open
source implementation of this method has been made available.
\end{abstract}

\ocis{(140.3430) Laser Theory; (140.3945) Microcavities; (140.5960) Semiconductor lasers.} 



\section{Introduction}

There has been great recent interest in using spatially nonuniform pumping of a laser to control the number and 
choice of lasing mode(s) \cite{bachelard_taming_2012,hisch_pump-controlled_2013,ge_enhancement_2014,liew_active_2014,liew_pump-controlled_2015,ge_selective_2015,ge_condensation_2016}.  
While previous works have involved both empirical approaches and those guided
by theory and simulations, the latter currently have drawbacks which limit their convenience and applicability
to detailed optimization studies.
Historically, the design and optimization of a laser begins and ends with design of the laser cavity. 
This approach has been used out of both convenience and necessity; many simulation tools can easily calculate the
eigenmodes of a passive cavity, but it is difficult to generalize these tools to include the effects of
the gain medium, such as mode competition, gain saturation and spatial hole-burning. For example, auxiliary equations can be added
to the finite difference time domain (FDTD) method to account for these effects, but these
add significant computational overhead \cite{bidegaray03,ho06,boehringer_part1_2008,huang_dynamical_2008,cerjan_noise_2015}. 
Frequency domain techniques, such as the steady-state \textit{ab initio} laser theory (SALT), can
provide an exact treatment of both the cavity geometry and gain medium, but no public implementation
is available \cite{tureci06,ge10,cerjan_csalt_2015}.
In lieu of a full treatment of the gain medium, the highest $Q$ eigenmodes of the passive cavity are assumed to be the lasing
modes, and in practice this ansatz works reasonably well provided that the cavity is uniformly pumped.
However, lasing modes are a function of both the cavity geometry and the spatial gain distribution within the cavity, 
and the laser properties can be significantly altered in the presence of non-uniform pumping.
Recent experimental and theoretical results have shown that optimizing the spatial pump profile can 
provide additional degrees of freedom, enabling the selection of individual lasing modes \cite{bachelard_taming_2012,ge_enhancement_2014,liew_active_2014,ge_selective_2015}, 
increasing the slope efficiency \cite{liew_pump-controlled_2015}, adding functionality by enabling tunable behavior \cite{hisch_pump-controlled_2013}, 
or increasing significantly the number of lasing modes at a given pump value \cite{ge_condensation_2016}.
While these works illustrate the potential impact of using non-uniform pumping to optimize the 
laser behavior, optimizing the spatial pump distribution experimentally is time consuming and the 
performance gain can be difficult to predict beforehand.

In this work, we present an analytic method and associated simulation tool which 
enables non-uniform pumping to be incorporated into laser design from the outset.
Specifically, we adapt an analytic approximation to SALT (SPA-SALT, the single pole approximation) \cite{ge10} so that it
can use the passive cavity resonances as an input. The passive cavity resonances can then be calculated using existing 
electromagnetic field simulation tools (e.g.~FDTD, FEM), 
after which SPA-SALT solves for the lasing thresholds and modal intensities 
as a function of the spatial pump distribution while  accurately accounting for the effects of mode competition, 
gain saturation and spatial hole-burning. This approach is valid when line pulling effects (including frequency locking \cite{sunada_stable_2013}) are negligible, such as for most semiconductor microcavity lasers,
as well as other laser systems with relatively high-Q passive cavities.
It provides a computationally efficient method for including the effects of 
non-uniform pump distributions in laser design. To illustrate breadth of this technique, 
we use this algorithm to optimize the spatial pump profile in a multi-mode chaotic cavity laser to achieve either 
highly multi-mode lasing, or single mode lasing in the same cavity. Furthermore, we confirm the validity
of this approximation through comparison with full SALT simulations. Finally, an open source implementation of 
this resonance SPA-SALT technique based on COMSOL is available in Code File 1 (Ref.~\cite{github_comsol}).

The remainder of this paper is organized as follows, in Sec.~\ref{sec:theory} we
provide the derivation of the resonance SPA-SALT equations. Section \ref{sec:results1} shows
semi-quantitative agreement between resonance SPA-SALT and exact simulations of the above threshold lasing behavior
calculated using SALT, demonstrating that
the passive cavity formulation of SPA-SALT is sufficient for guiding device design.
Section~\ref{sec:results2} reports on enhancing the multi-mode
behavior of cavities through numerical non-linear optimization of the non-uniform pump profile.
In Sec.~\ref{sec:results3}, we then show how changing the pump profile can be used to suppress
multi-mode behavior, again finding semi-quantitative agreement between SPA-SALT and SALT,
and demonstrating that single-mode behavior can be promoted through optimization
of the pump profile.
Finally, Sec.~\ref{sec:conc}
will offer some concluding remarks.

\section{Passive single pole approximation \label{sec:theory}}

As the SPA-SALT equations are based on SALT, which provides a quite general and accurate theory of multimode steady
state lasing, it is useful to provide a brief overview of that theory.
SALT was first derived nearly a decade ago as a frequency-domain formulation of the 
semi-classical Maxwell-Bloch equations in steady-state, which can be solved efficiently to provide
a nearly exact description of the spatial degrees of freedom of the laser system\cite{tureci06,ge10,cerjan_csalt_2015}.
SALT has two features that distinguish it from previous approaches to solving the Maxwell-Bloch equations,
the ability to treat cavities with an arbitrary geometry, such as chaotic or random lasers \cite{nockel97,gmachl98,cao_random_1999,cao05},
and the capability to describe essentially exactly the space-dependent saturation of the gain
medium (spatial hole-burning). The fundamental SALT equations are coupled non-linear wave equations
for each active lasing mode, which take the form:
\begin{align}
\left[ \left(\boldsymbol{\nabla} \times \boldsymbol{\nabla} \times \right) - \left(\varepsilon_c(\mathbf{x}) + \chi_g(\mathbf{x},\omega_\mu,D_0)\right) \frac{\omega_\mu^2}{c^2} \right] \boldsymbol{\Psi}_\mu(\mathbf{x}) = 0, \;\;\; \mathbf{x} \in C, \notag \\
\left[ \left(\boldsymbol{\nabla} \times \boldsymbol{\nabla} \times \right) - n_0^2 \frac{\omega_\mu^2}{c^2} \right] \boldsymbol{\Psi}_\mu(\mathbf{x}) = 0. \;\;\; \mathbf{x} \notin C, \label{eq:LaseWave}
\end{align}
The $N_L$ lasing modes interact through the non-linear space-dependent saturation of the gain medium, which
for a two-level Bloch atomic system is written as
\begin{equation}
\chi_g(\mathbf{x},\omega,D_0) = \frac{\gamma_\perp D_0(\mathbf{x})}{\omega - \omega_a + i\gamma_\perp}\left(\frac{1}{1 + \sum_\nu^{N_L} \Gamma_\nu |\boldsymbol{\Psi}_\nu|^2}\right). \label{eq:chi}
\end{equation}
In these equations, $\boldsymbol{\Psi}_\mu$ and $\omega_\mu$ are the spatial profile and frequency
of the lasing modes, $c$ is the speed of light in vacuum, $\varepsilon_c$ is the dielectric function of the passive cavity, which 
is contained in a spatial domain denoted by $C$, $n_0$ is the index of refraction outside
the cavity, $D_0(x)$ is the pump strength, $\omega_a$ is the atomic transition frequency, $\gamma_\perp$ is the gain width (polarization dephasing rate),
$\Gamma_\nu$ is the Lorentzian gain factor $\Gamma_\nu = \gamma_\perp^2/( (\omega_\nu-\omega_a)^2 + \gamma_\perp^2)$,
and the pump strength and fields have been written in normalized units \cite{tureci06,ge10}.
As lasing modes are self-generated within the cavity, they must satisfy a purely radiating 
(Sommerfeld) boundary condition,
\begin{equation}
\lim_{r \to \infty} r^{\frac{d-1}{2}} \left( c\partial_r - i \omega_\mu \right) \boldsymbol{\Psi}_\mu(\mathbf{x}) = 0, \label{eq:SRC}
\end{equation}
in which $r\equiv|\mathbf{x}|$, and $d$ is the dimensionality of the system \cite{sommerfeld}.
Enforcing this purely outgoing boundary condition upon Eq.~(\ref{eq:LaseWave}) results in a quantization condition
on the allowed frequencies of the lasing modes. Taken together, Eqs.~(\ref{eq:LaseWave})-(\ref{eq:SRC}), comprise
the SALT equations, $N_L$ coupled differential equations, one for each active lasing mode, where $N_L$ is determined 
self-consistently as the pump strength, $D_0$, is varied.  The SALT equations are usually
solved by introducing a complete set of constant flux (CF) states at every outgoing frequency by making use of
the quantization provided by the outgoing boundary condition \cite{ge10}, however direct solution methods are also 
possible \cite{esterhazy14}. The SPA-SALT equations used here are most easily derived from the CF-state formulation \cite{ge10}.

Unfortunately, the explicit numerical implementation of the outgoing boundary condition, Eq.~(\ref{eq:SRC}),
can be challenging, especially in multiple dimensions, as it is frequency dependent. 
Outgoing boundary conditions are usually implemented numerically
in an implicit manner, without directly specifying the outgoing frequency, 
through the use of an impedance matched absorbing region surrounding
the finite simulation domain \cite{taflove}. This inability to specify an arbitrary outgoing
frequency makes it difficult for these implicit methods to solve for the basis of CF states needed for the SPA-SALT approximation. 
Instead, without adding gain to the cavity, using such an implicit outgoing boundary condition
naturally yields the passive cavity resonances (as a subset of all of the modes of the finite system); the resonances satisfy
\begin{equation}
\left[ \left(\boldsymbol{\nabla} \times \boldsymbol{\nabla} \times \right) - \left(\varepsilon_c(\mathbf{x}) + \varepsilon_{abs}(\mathbf{x}) \right)\frac{\tilde{\omega}_m^2}{c^2} \right] \boldsymbol{\varphi}_m(\mathbf{x}) = 0, \label{eq:PCM} \\
\end{equation}
where $\tilde{\omega}_m \in \mathbb{C}$ is the (complex) frequency of the passive cavity resonance,
and $\varepsilon_{abs}(\mathbf{x})$ is the impedance-matched absorbing boundary layer at the edges
of the simulation domian.
The pairs $\{ \tilde{\omega}_m, \boldsymbol{\varphi}_m \}$ constitute a countably infinite
set, quantized by the Dirichlet boundary condition at the edge of the finite simulation domain. 
The difference between the threshold lasing modes, which can each be expressed as a single CF state, and the passive cavity resonances is critical
when considering line pulling effects from the gain medium, as the lasing frequency is not necessarily 
given by the real part of the passive cavity resonance. This distinction is especially pronounced
in low-$Q$ ``bad cavity'' systems such as random lasers, where the width of the gain curve can be much less than the decay rate of the cavity, 
$\gamma_\perp \ll \gamma_c = -2\im[\tilde \omega_m]$ \cite{goetschy_random_laser}. However, for most traditional laser systems, 
$\gamma_c \ll \gamma_\perp$, which results in negligible line-pulling effects, and thus the
frequencies and spatial profiles inside the cavity of the passive cavity resonances and the relevant active cavity CF states are nearly identical \cite{tureci06}.

The single pole approximation was derived as a simplified form of the SALT equations, and makes use of the
fact that the non-interacting lasing modes at threshold are each given by exactly one CF state at
the lasing frequency \cite{tureci06,ge10}. 
Here, we exploit this similarity between the dominant CF state for each lasing mode and the passive cavity resonances 
to derive the SPA-SALT equations in terms of
the passive cavity mode frequencies and spatial profiles, letting $\boldsymbol{\Psi}_\mu = a_\mu \boldsymbol{\varphi}_\mu$.  
The derivation of the resonance version of SPA-SALT is very similar to the derivations using CF-states  \cite{ge10}, so we will only highlight
the major differences here, and provide all of the equations necessary for numerical solution of the problem.

Assuming that we have used a separate numerical package such as COMSOL Multiphysics to solve for the
frequencies and spatial profiles of the passive cavity resonances, we renormalize the modes over the
volume of the cavity, rather than the entire simulation domain, as
\begin{equation}
\int_C \varepsilon_c(\mathbf{x}) \boldsymbol{\varphi}_\mu(\mathbf{x}) \boldsymbol{\varphi}_\mu(\mathbf{x}) d\mathbf{x} = 1.
\end{equation}
The passive cavity frequencies and spatial profiles can then be directly used in Eqs.~(\ref{eq:LaseWave}) and (\ref{eq:PCM}) to solve for the non-interacting lasing
mode thresholds as,
\begin{equation}
D^{(\mu)} = \left| \left(\frac{\re[\tilde{\omega}_\mu] - \omega_a + i\gamma_\perp}{\gamma_\perp}\right) \left(\frac{\tilde{\omega}_\mu^2 - \re[\tilde{\omega}_\mu]^2}{\re[\tilde{\omega}_\mu]^2}\right) \right| \left(\frac{1}{\bar{f}_\mu}\right), \label{eq:nonint}
\end{equation}
in which the the pump overlap integral is given by
\begin{equation}
\bar{f}_\mu = \left|\int_C f(\mathbf{x}) \boldsymbol{\varphi}_\mu^2(\mathbf{x}) d\mathbf{x} \right|, \label{eq:f}
\end{equation}
such that $\int_C f(\mathbf{x}) d\mathbf{x} = V$, the volume of the cavity. Note that the pump
overlap integral did not appear in the original formulation of SPA-SALT, where it was instead included
in the solution to, and normalization of the CF states respectively.

The SPA-SALT approximation assumes the lasing modes and frequencies are known and need not be obtained self-consistently
(in the resonance approximation they are determined by Eq.~(\ref{eq:PCM})).
With this assumption 
Eqs.~(\ref{eq:LaseWave}) can be inverted as in \cite{ge10}, to yield
\begin{equation}
\frac{D_0}{D_0^{(\mu)}} - 1 = \sum_\nu \Gamma_\nu I_\nu \chi_{\mu \nu}, \label{eq:SPASALT}
\end{equation}
in which $D_0^{(\mu)},I_\mu = |a_\mu|^2$ are respectively the non-interacting lasing threshold and intensity of the $\mu$th lasing mode,
constrained to be positive.  The modal overlap is given by
\begin{equation}
\chi_{\mu \nu} = \left| \int_C \boldsymbol{\varphi}_\mu(\mathbf{x}) \varepsilon_c(\mathbf{x})  \boldsymbol{\varphi}_\mu(\mathbf{x}) |\boldsymbol{\varphi}_\nu(\mathbf{x})|^2 d\mathbf{x}\right|.
\end{equation}
Eq.~(\ref{eq:SPASALT}) combined with Eq.~(\ref{eq:PCM}) are the fundamental equations of resonance SPA-SALT, which
yield the modal lasing thresholds, intensities and power-slopes above threshold (the modal frequencies are assumed to be
the real part of the passive cavity resonances). These equations taking into account the effects of gain competition through the
interaction coefficients $\chi_{\mu \nu}$, and the effects of a spatially varying pump through its appearance in the non-interacting
threshold equation, (\ref{eq:nonint}).

Despite the relative simplicity of Eq.~(\ref{eq:SPASALT}), it does not immediately specify
the number of active lasing modes for any arbitrary pump value, and is instead dependent
upon the non-interacting threshold pump values. The interacting threshold pump values are determined
from their respective non-interacting values as,
\begin{equation}
D_{int}^{(\mu)} = \frac{1}{1-\lambda_\mu}D^{(\mu)}, \label{eq:int1}
\end{equation}
in which $\lambda_\mu \in [0,1]$ is the generalized mode competition parameter defined in \cite{ge10}, 
\begin{equation}
\lambda_\mu = \frac{\sum_{\nu=1}^{N_L} A_{\mu\nu}(c_\nu D^{(\mu)} - b_\nu)}{1 - \sum_{\nu=1}^{N_L} A_{\mu\nu} b_\nu}, \label{eq:int2}
\end{equation}
where we have used the definitions,
\begin{gather}
A_{\mu\nu} = \Gamma_\nu \chi_{\mu\nu}, \\
b_\mu = \sum_{\nu=1}^{N_L} \left(A^{-1}\right)_{\mu\nu}, \\
c_\mu = \sum_{\nu=1}^{N_L} \frac{\left(A^{-1}\right)_{\mu\nu}}{D^{(\nu)}}.
\end{gather}
In the absence
of mode competition, $\lambda_\mu = 0$, and the interacting thresholds are identical to their
non-interacting counterparts, while $\lambda_\mu = 1$ indicates gain clamping, where the $\mu$th
mode never reaches threshold. Finally, after calculating the interacting mode thresholds using Eqs.~(\ref{eq:int1}) and (\ref{eq:int2}),
one can solve for the modal intensities of the lasing modes at any pump value as,
\begin{equation}
I_\mu = c_\mu D_0 - b_\mu. \label{eq:inten}
\end{equation}
Thus, using Eqs.~(\ref{eq:int1}) and (\ref{eq:inten}), the above-threshold interacting semi-classical
properties of lasers can be calculated from the knowledge of the passive cavity resonances
with little additional computational effort. 

Resonance SPA-SALT provides a convenient tool for studying any kind of cavity with a complex geometry 
for which the resonances can be efficiently calculated numerically, e.g. photonic crystal lasers or various kinds
and shapes of microcavity lasers.  Once the resonances are known, the effect of the spatial distribution of the 
pump can be included with negligible computational effort through changing the parameter ${\bar{f}_\mu}$ in 
Eq.~(\ref{eq:nonint}).  Optimization of specific laser properties via non-uniform pumping can then be modeled.
In the next section we will apply this formalism to a specific case, the optimization of a chaotic cavity laser for highly multimode
operation; to validate the method we will compare the results of resonance SPA-SALT with full SALT simulations.

\section{Results}

\subsection{Optimizing multimode behavior through passive cavity design\label{sec:results1}}

\begin{figure}[t!]
\centering
\includegraphics[width=0.95\textwidth]{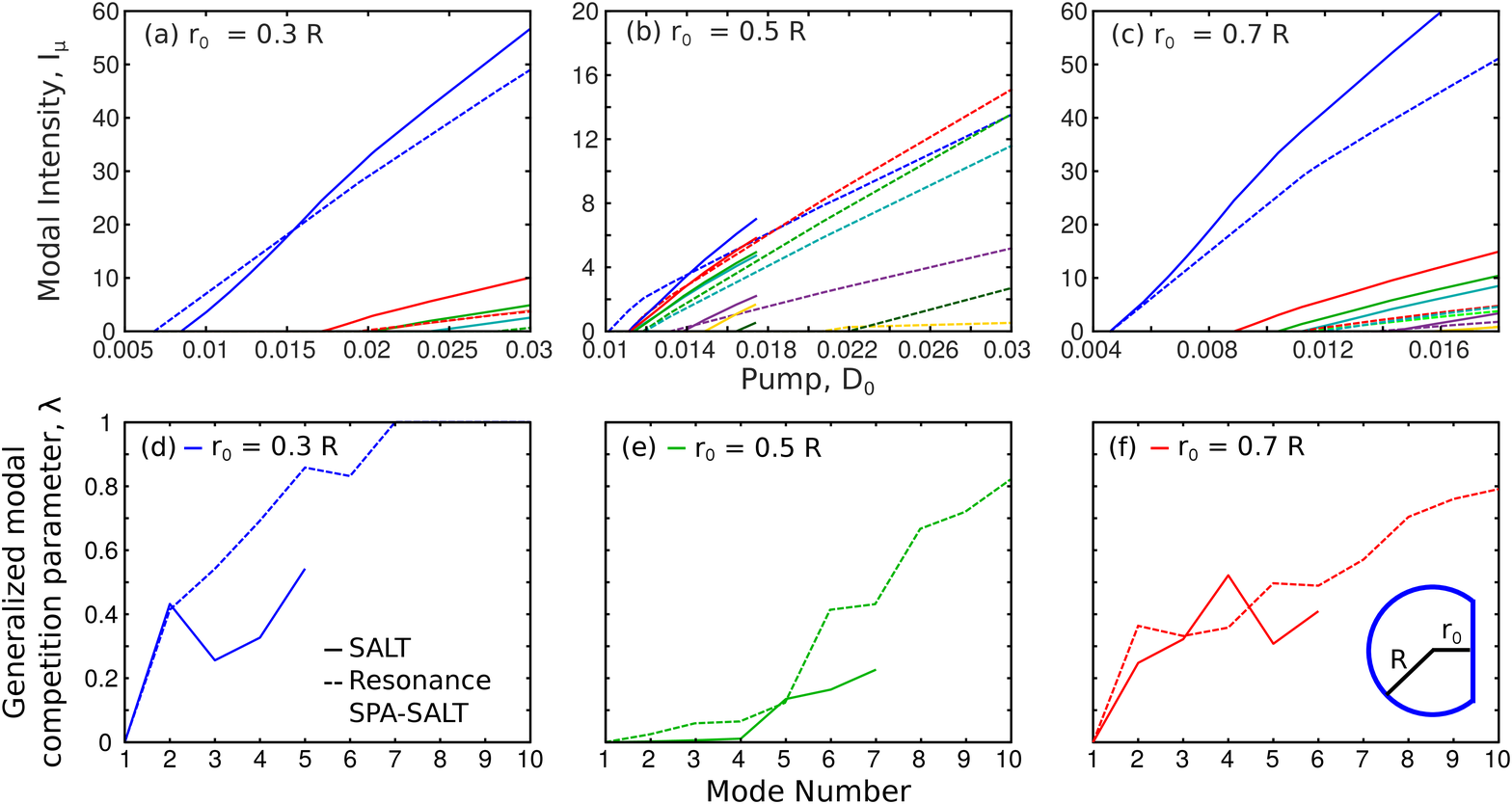}
\caption{
(a-c) Plot of the modal intensity $I_\mu$ as a function of the pump parameter $D_0$ 
for D-shaped cavities with $R = 5\mu$m, $n=3.5$, filled with a gain medium with a central wavelength of $\lambda_a = 1\mu$m, and width $\gamma_\perp = 30$nm. 
Different choices of $r_0$ are shown in each plot, (a) $r_0 = 0.3R$, (b) $r_0 = 0.5R$, and (c) $r_0 = 0.7R$.
Solid lines show the direct calculation using full SALT simulations, while dashed lines show 
resonance SPA-SALT results. The SALT simulations were
performed using 100 CF basis states, and were run to $\sim 6$ lasing modes.
In (b), the SALT simulations terminate after seven lasing modes when the simulations became
too memory intensive to continue.
(d-f) Plots of the generalized modal competition parameter, $\lambda$, given in Eq.~(\ref{eq:int2}),
for the same cavity.
Different choices of $r_0$ are shown in each plot, (d) $r_0 = 0.3R$, (e) $r_0 = 0.5R$, and (f) $r_0 = 0.7R$. \label{fig:spasalt}}
\end{figure}

One way to design a low spatial coherence light source is to construct a laser
with many independent lasing modes \cite{redding_spatial_2011,redding_speckle-free_2012,redding_dcav_pnas_2015}.
If the cavity is sufficiently complex, each lasing mode
has a distinct spatial field pattern, and thus adding the emitted fields of many modes leads
to a reduction in the spatial coherence. As an initial comparison of the passive cavity
formulation of SPA-SALT with SALT, we choose to study
chaotic ``D-shaped'' cavities which have recently been shown experimentally to generate
laser emission with low spatial coherence \cite{redding_dcav_pnas_2015}. D-shaped cavities consist
of a disk with radius $R$, with a section removed along a chord parameterized by
$r_0$, as shown in the inset of Fig.~\ref{fig:spasalt}. These cavities are known to support chaotic
ray dynamics \cite{bunimovich_ergodic_1979,ree_classical_1999}, such that 
if one neglects the out-coupling losses and considers the system as an `ideal billiard,' 
generic ray orbits for the D-shaped cavity cover ergodically the entire area of the cavity.

Broadly speaking, there are three requirements that must be satisfied in the design of a laser cavity suitable as
an incoherent light source. First, there must be a large density of passive cavity modes with similar $Q$-factors close to the atomic resonance, 
such that the device has the potential for many lasing modes. Second, the lasing modes must have distinct 
transverse fields patterns so that the total field tend to self-average (longitudinal modes of a linear cavity do not have
this property). Third, the mode-competition between these modes needs to be minimized,
so that all of these modes are able to reach threshold even in the
presence of other active modes. This is why random lasers and degenerate lasers are good candidates
for producing low coherence light sources, they have many modes
with similar $Q$-factors, whose spatial profiles are uniformly distributed
throughout the cavity, effectively minimizing gain competition \cite{redding_spatial_2011,redding_speckle-free_2012,nixon_efficient_2013,hokr_single-shot_2014,hokr_narrow-band_2015}. For comparison,
perfectly circular disk lasers are a poor choice for a low-coherence laser. While disk cavities
do support many high-$Q$ modes, these are all whispering gallery modes and are localized
at the edge of the cavity, which leads to large mode competition, such that only a few of these
modes reach threshold. 

The chaotic ray trajectories of the D-shaped cavities results in passive cavity resonances
whose spatial profiles are relatively uniformly distributed across the cavity, leading to a narrow
distribution of $Q$ values and no very high Q modes with special geometry.  Nonetheless, the 
high index mismatch at the boundary is sufficient to bring the lasing threshold into a reasonable range.
Thus, optimizing multi-mode behavior in
such a cavity depends upon reducing the effects of mode competition, which
can be calculated exactly using SALT, and estimated with resonance SPA-SALT.
In Fig.~(\ref{fig:spasalt})
we calculate the modal intensities (a)-(c), and generalized modal competition parameters, $\lambda$ (d)-(f),
using the passive cavity mode profiles compare this to
full SALT simulations of the same cavity. Semi-quantitative agreement is seen between these
two different computational methods, and both calculations independently agree on the
geometry of the D-shaped cavity, $r_0 = .5R$, which minimizes the effects of mode competition, and thus
produces the most lasing modes for similar values of the pump. However,
full SALT simulations of the cavity in this multi-mode regime require hundreds of hours of
computational time to both generate the necessary CF basis states and solve the coupled non-linear
equations above threshold, while the resonance SPA-SALT calculations require mere minutes
after calculating the passive cavity resonances.

\subsection{Enhancing multi-mode behavior with non-uniform pumping \label{sec:results2}}

\begin{figure}[t!]
\centering
\includegraphics[width=0.50\textwidth]{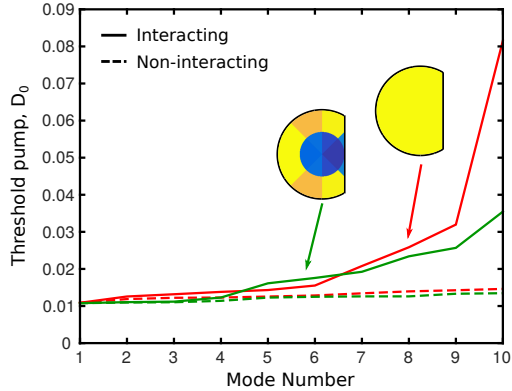}
\caption{Plot of the non-interacting (dashed lines) and interacting (solid lines) modal thresholds
for the first $10$ lasing modes for a D-shaped cavity with $R = 5\mu$m, $n=3.5$, filled with a gain medium with a central wavelength of $\lambda_a = 1\mu$m, and width $\gamma_\perp = 10$nm
using different pump profiles calculated using SPA-SALT with the passive cavity resonances
generated by COMSOL Multiphysics. The thresholds for the uniform pump are shown in red and
those generated using a genetic algorithm to minimize Eq.~(\ref{eq:fit1}) are in green. 
For the inhomogeneous pumping scheme, there are $12$ independent electrical contacts which divide
the cavity radially in two, and angularly in eight.\label{fig:adap}}
\end{figure}

Having now demonstrated that resonance SPA-SALT agrees semi-quantitatively with the exact theory, one can use SPA-SALT
as part of an optimization scheme to guide laser design. In this section, we explore the effects
of non-uniform pumping on chaotic cavities to further increase their multi-mode behavior,
and demonstrate that resonance SPA-SALT is compatible with existing non-linear optimization
methods.

To begin, we consider a D-shaped cavity with independently tunable electrical contacts providing
the current to the gain medium $\{p_i\}$, to provide a spatially inhomogeneous gain profile for Eq.~(\ref{eq:f}).
(We assume there is no significant gain diffusion in the medium).
These electrical contacts are segmented both angularly and radially,
but together cover the entire cavity. 
Most non-linear optimization techniques work to minimize a fitness function with
respect to an array of inputs, and so to optimize a pump profile for $N_L$ lasing modes
we choose the fitness function, $F$, to be
\begin{equation}
F(\{p_i\}) =  \prod_{n=1}^{N_L} \frac{D_{adap,int}^{(n)}}{D_{uni}^{(1)}}, \label{eq:fit1} 
\end{equation}
where $D_{uni}^{(1)}$ is the threshold of the first lasing mode with uniform pumping, and the
interacting thresholds, $D_{adap,int}^{(n)}$, with non-uniform pumping are listed in ascending
order. The normalization of $1/D_{uni}^{(1)}$ is chosen for numerical stability.
Using this fitness function, non-uniform pumping is found to enhance the multi-mode behavior
of chaotic cavities, as shown in Fig.~\ref{fig:adap} in green. Here, we have used a genetic algorithm
to perform this optimization, and see that the total pump power required to achieve
$N_L=10$ lasing modes can be reduced by over a factor of $2$.

\subsection{Reducing multi-mode behavior with non-uniform pumping \label{sec:results3}}

\begin{figure}[t!]
\centering
\includegraphics[width=0.95\textwidth]{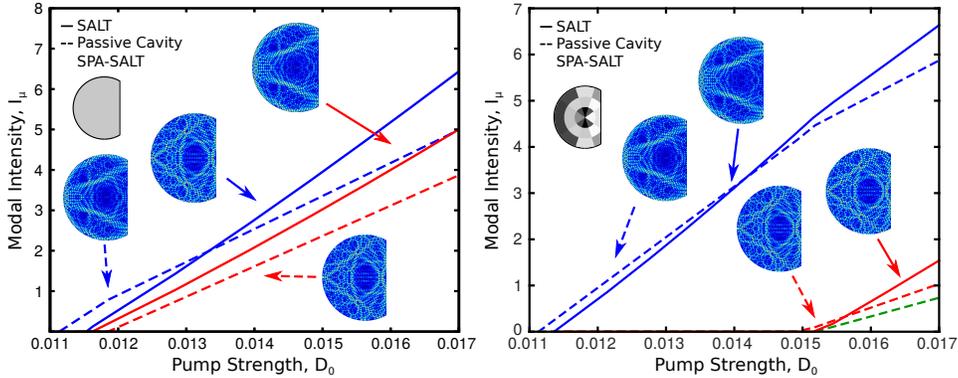}
\caption{Plots of the modal intensities, $I_\mu$, given in Eq.~(\ref{eq:inten}),
for a D-shaped cavity with $R = 4\mu$m, $r_0 = 2\mu$m, $n=3.5$, filled with a gain medium with a central wavelength of $\lambda_a = 1\mu$m, and width $\gamma_\perp = 10$nm. 
Solid lines show the direct calculation using full SALT simulations, while dashed lines show the SPA-SALT results
generated using passive cavity information generated by COMSOL Multiphysics. (Left panel) Uniformly pumped D-shaped cavity,
as indicated in the inset. (Right panel) Inhomogeneously pumped D-shaped cavity, as indicated in the inset. Here,
the pump profile has been optimized using Eq.~(\ref{eq:fit2}) to increase the range
of single mode operation while maintaining the same threshold pump as the uniformly pumped cavity. There are $23$ electrical
contacts here, which divide the cavity radially in three, and angularly in ten. \label{fig:reduce1}}
\end{figure}

To demonstrate the flexibility of the pump optimization method for a given system, we now
change our desired target for optimization, and try to inhibit multimode lasing through non-uniform
pumping of a similar D-shaped cavity.  This type of pump control has been demonstrated in previous work
 \cite{ge_enhancement_2014,liew_active_2014,liew_pump-controlled_2015,ge_selective_2015}, although 
not with this method and not in such a highly multimode cavity as this.
In Fig.~\ref{fig:reduce1}, we show the modal intensities as a function
of pump strength calculated using SALT (solid lines), and resonance SPA-SALT (dashed lines), 
where the passive cavity resonances have again been calculated using COMSOL Multiphysics. In the left panel of Fig.~\ref{fig:reduce1},
we show the results for uniform pumping, wherein the first two lasing modes are
seen to have similar thresholds, indicating that the effects of mode
competition are weak.

However, as shown in the right panel of Fig.~\ref{fig:reduce1}, by changing the pump profile of the cavity
to have the same overall gain but an inhomogeneous distribution,
the effects of mode competition can be increased, resulting in a substantially
increased range of single mode operation. Here we have chosen to optimize
the pump profile to promote single mode lasing while maintaining the same
lasing threshold to facilitate switching between single-mode and multi-mode behavior.
To do so, we are instead using a fitness function to both select for the same lasing threshold as the uniformly
pumped cavity, while increasing the interacting threshold of the second mode relative to the first,
\begin{equation}
F(\{p_i\}) =  \frac{D_{adap}^{(1)}}{D_{adap,int}^{(2)}} + \left| 1-\frac{D_{adap}^{(1)}}{D_{uni}^{(1)}} \right|. \label{eq:fit2}
\end{equation}
Thus, for the same cavity as is shown in the left panel of Fig.~\ref{fig:reduce1}, the
regime of single mode operation can be increased by a factor of $4$.
In both simulations, we see semi-quantitative agreement
between SALT and resonance SPA-SALT for all of the important quantities, the first lasing
threshold, the interacting second lasing threshold, the power slopes, and the full
spatial profiles of the modes within the cavity, though in the case of uniform pumping,
the spatial profiles of the first and second modes are switched between the two computational
methods. Again, the full SALT simulations are found to require substantially increased
computational time when compared against the resonance SPA-SALT method.

\section{Summary \label{sec:conc}}

In conclusion, we have demonstrated that non-linear properties of above threshold
lasers with $Q > 100$ can be calculated using knowledge of the
passive cavity resonances and the resonance-based SPA-SALT equations. These results show semi-quantitative agreement
with the full SALT simulations, for all of the semi-classical properties of lasers of interest,
non-interacting and interacting modal thresholds, power slopes, and mode profiles.
Using resonance SPA-SALT, we were then able to increase the multi-mode behavior of chaotic D-shaped
cavities, demonstrating a potential role for selective pumping in the development of improved incoherent laser light sources.
We note that, due to fabrication imperfections, often the goal of device optimization methods
is to inform the experimental design process, not necessarily to achieve a highly accurate quantitative 
prediction for results. For the purpose of this type of exploratory device design, resonance SPA-SALT
is able to provide the same insights as full SALT simulations, but using widely
available software for resonance calculations in conjunction with an open source implementation of the method
derived here.


\section*{Funding}
A.C. and A.D.S. acknowledge the funding support by the National Science Foundation (NSF) (DMR-1307632).
S.F.L. and H.C. acknowledge the funding support by the National Science Foundation (NSF) (ECCS1509361) 
and the Office of Naval Research (ONR) (MURI N00014-13-1-0649).

\end{document}